\documentclass[twocolumn,showpacs,preprintnumbers,amsmath,amssymb]{revtex4}
\usepackage{graphicx}
\usepackage{dcolumn}
\usepackage{bm}
\usepackage{hyperref}

\begin{document}
\preprint{}
\title{Boundary conditions in first order gravity: Hamiltonian and Ensemble}
\author{Rodrigo Aros}
\affiliation{Departamento de Ciencias F\'{\i}sicas\\ Universidad Andr\'es Bello, Republica 252,
Santiago,Chile}
\date{\today}
\begin{abstract}
In this work two different boundary conditions for first order gravity, corresponding to a null
and a negative cosmological constant respectively, are studied. Both boundary conditions allows to
obtain the standard black hole thermodynamics. Furthermore both boundary conditions define a
canonical ensemble. Additionally the quasilocal energy definition is obtained for the null
cosmological constant case.

\end{abstract}

\maketitle

\section{Introduction}

It is well known that the variation of the action on shell, provided the boundary conditions, must
vanish in order to have a well defined variational principle, and that simultaneously the same
boundary conditions must allow the existence of solutions for the equations of motion.

However there is a third role, which arises from the formal connection between quantum field
theory and statical mechanics, the boundary conditions define the ensemble in the statistical
mechanical counterpart. For instance the partition function in the canonical ensemble can be
written as the path ordered integral
\begin{equation}\label{PartitionFunctionFeynman}
Z(\beta) =  \oint \mathcal{D} x e^{i I|_{\tau=-i\beta}},
\end{equation}
provided $I$ effectively be such that the temperature $\beta^{-1}$, said the inverse of the
period, be fixed. In principle there is a suitable action for any other ensemble.

For gravity the same connection scenario, at least at tree level, seems to exist
\cite{Hawking:1983dh}, justifying a general analysis. To study gravity in this direction a
Hamiltonian analysis \cite{Regge:1974zd,Brown:1993bq,Peldan:1994hi} can be very useful since one
can expect that the Hamiltonian charges be related with the mass, angular momentum and entropy in
the statistical mechanics side.

Unfortunately gravity in asymptotically flat spaces is not a well defined statistical mechanics
system. Actually it is necessary to \textit{put the system in a box} to perform computations
\cite{Brown:1993bq}\footnote{To have an idea of the reasons see appendix \ref{Naive}.}.
Alternatively this can be usually regularized by introducing a background configuration.

On the other hand, gravity with a negative cosmological constant, at least formally, is a well
defined statistical mechanical system \cite{Hawking:1983dh}. This can be foreseen since a negative
cosmological constant introduces a negative pressure which constrains the fields producing,
roughly speaking, the effect of a box. Remarkably the most of results of asymptotically flat case
usually can be obtained through the limit $\Lambda \rightarrow 0$, leading to consider $\Lambda$
as a regulator of the theory of gravity.

In this work both $\Lambda=0$ and $\Lambda<0$ will be addressed, however with different
approaches.

\subsection{First Order gravity}

Fermions represents a different scenario in gravity. Fermions can not be directly incorporated in
metric gravity because, roughly speaking, the group of diffeomorphism does not have half integer
representations. Fermions must be represented by spinors, and to incorporate spinors is necessary
to introduce a local Lorentz group where they can be realized. This can be done by introducing a
local orthonormal basis for the co-tangent space, called vielbein. The vielbein is usually written
in terms of the set of differential forms $e^{a}=e^{a}_{\mu} dx^{\mu}$. The metric here is the
composed field $g_{\mu\nu}=e^{a}_{\mu}e^{b}_{\nu}\eta_{ab}$. In four dimensions one usually speaks
of a vierbein instead.

The introduction of the vielbein motivates a reformulation of gravity
\cite{Zumino:1986dp,Regge:1986nz} where the corresponding Lorentz connection is an additional
independent field. The Lorentz connection is called the spin connection and also is written in
terms of the differential forms $\omega^{ab}=\omega^{ab}_{\mu}dx^{\mu}$. Although it is direct to
confirm that this reformulation is essentially different in many aspects most of the results of
metric gravity in this reformulation are recovered. This work aims to analyze one those aspects.
This new formulation is usually called first order gravity.

The four dimensional Einstein Hilbert action in first order formalism reads
\begin{equation}\label{EHaction}
    I_{EH} = \int_{\mathcal{M}} R^{ab}\wedge e^{c}\wedge e^{d}\varepsilon_{abcd},
\end{equation}
where \[R^{ab}=d\omega^{ab} + \omega^{a}_{\hspace{1ex} c} \wedge \omega^{cb}=
\frac{1}{2}R^{ab}_{\hspace{2ex}cd}e^{c}\wedge e^{d},\] being $R^{ab}_{\hspace{2ex} cd}$ the
Riemann tensor. $\varepsilon_{abcd}=\pm 1,0$ stands for the complete antisymmetric symbol
\footnote{antisymmetric symbol differs form Levi-Civita pseudo tensor by $\sqrt{g^{(4)}}$. To
avoid confusion in this work only antisymmetric symbols will be used and any determinant of the
metrics will be written explicitly}.

The variation of Eq.(\ref{EHaction}) yields the two set of equations of motion
\begin{eqnarray}
  \delta e^{d} &\rightarrow& R^{ab}\wedge e^{c}\varepsilon_{abcd}=0 \label{EinsteinEquations},\\
  \delta \omega^{ab} &\rightarrow& T^{c}\wedge e^{d}\varepsilon_{abcd}=0\label{TorsionEquations},
\end{eqnarray}
where $T^{a}=de^{a} + \omega^{a}_{\hspace{1ex} b}\wedge
e^{b}=\frac{1}{2}T^{a}_{\hspace{1ex}bc}e^{b}\wedge e^{c}$ corresponds to the torsion two form with
$T^{a}_{\hspace{1ex}bc}$ the torsion tensor. Note that Eq.(\ref{TorsionEquations}) is an algebraic
equation, with solution $T^{a}_{\hspace{1ex} cd}=0$. Once this is replaced on
Eq.(\ref{EinsteinEquations}) they become the Einstein equations. Thus any solution of the metric
formalism is recovered on-shell by this formulation.

If fermions are presented $T^{a}\neq 0$, for instance in the presence of gravitinos
\[
 T^{a} \sim  \bar{\Psi}\gamma^{a}\Psi,
\]
and thus first order in this case presents a different kind of solutions.

In general since $e^{a}$ and $\omega^{ab}$ are independent fields one could expect that there were
an independent conjugate momentum for each one. However in four dimensions the conjugate of
momentum of $e^{a}$ is contained in $\omega^{ab}$ or viceversa.
\cite{Banados:1997hs,Contreras:1999je}. This leads to the definitions of two equivalent phase
spaces that can be mapped into each other readily. One can even confirm the equivalence of their
path order integrals \cite{Contreras:1999je}. These equivalent phase spaces are called $e$ and
$\omega$-frames respectively.

One remarkable result of first order gravity is to reproduce the path ordered integral of metric
formalism. Once its momenta are integrated out the resulting expression is the same obtained in
the metric formalism once its corresponding momenta, usually denoted $\pi^{ij}$, are integrated
out \cite{Aros:2003bi}, \textit{i.e}
\[
 \int De^{a}_{i} D\pi_{a}^{i}\, e^{I_{Ham}} \equiv \int Dg_{ij}\, e^{\int R\sqrt{g} d^{4}x}.
\]

However both results -first order and metric - are made ignoring the boundary terms. It will be
very interesting to address the same computation in first order gravity considering the presence
of those boundary terms. Results in the metric formalism considering the boundary terms are very
promising and for instance they are connected the entropy of black holes as observed in
\cite{Brown:1997rp}. In this case one can expect some deviation at first loop since one should sum
over also torsional degrees of freedom at the boundary.

\subsection{Energy}

The quest for a definition of energy in gravity has been addressed by many authors (see for
instance
\cite{Komar1959,Regge:1974zd,Gibbons:1979xm,Lee:1990nz,Wald:1997qz,Henneaux:1999ct,Aros:1999id,Barnich:2001jy,Barnich:2003xg}).
In principle a definition of energy, even for classical mechanics, relays on the boundary terms of
the action, so it is in the case of gravity. In general the boundary terms fix the ground state of
the system, but also the definition of a finite energy might relay on them as well (see for
instance \cite{Hawking:1996fd,Aros:1999id}).

The connection between boundary conditions and a definition the energy will be explored in this
work trying to shed some more light into the problem of energy in gravity. Here two definitions of
energy for four dimensional first order gravity will be used, each connected boundary conditions
for null and negative cosmological constants. However the two different boundary conditions will
be shown to recover the \textit{(grand) canonical ensemble}.

It is worth to mention another approach to this subject in \cite{Fatibene:2000jm}, where another
kind of first order gravity is discussed.

\subsection{The space}

The space to be discussed in this work corresponds to a topological cylinder. One can picture it
as $\mathcal{M}=\mathbb{R}\times \Sigma$ where $\Sigma$ corresponds to a 3-dimensional spacelike
hypersurface and $\mathbb{R}$ stands for the time direction and formally is a segment of the real
line. In this way the boundary of the space is given by $\partial\mathcal{M}=
\mathbb{R}\times\partial\Sigma\cup \Sigma_{+} \cup \Sigma_{-}$, where $\Sigma_{\pm}$ are the upper and
lower boundaries of the topological cylinder and $\partial \Sigma$ is the boundary of $\Sigma$.
Finally $\partial \Sigma$ represents at least the asymptotical spatial region of the manifold,
however in the case of a black hole be considered $\partial \Sigma=\partial \Sigma_{\infty} \oplus
\partial \Sigma_{H}$, where $\partial \Sigma_{H}$ stands for the horizon.

\section{$\Lambda =0$ a definition of energy and entropy}

To discuss this case one can begin by recalling that the phase space of four dimensional first
order gravity can be described in either the $e$-frame or the $\omega$-frame. For $\Lambda=0$ it
seems more suitable to work in the $e$-frame.

\subsection{Fixing the fields}
The variation of the EH action (\ref{EHaction}) yields the boundary term
\begin{equation}\label{Variation}
\delta I_{EH}|_{\textrm{on shell}} =  \int_{\mathcal{\partial M}} \delta \omega^{ab}\wedge
e^{c}\wedge e^{d}\varepsilon_{abcd},
\end{equation}
which implies that the EH action could be a proper action principle provided $\delta\omega^{ab}=0$
at the boundary. For reasons that would be clear later, instead of fixing the connection at
$\partial \mathcal{M}$ in this work the vierbein will be fixed. This is similar to the condition
$\delta g_{ij}|_{\partial \mathcal{M}}=0$ discussed in \cite{Brown:1992br}. This leads to modify
the action by adding the boundary term
\begin{equation}\label{omegaAB}
    -\int_{\partial \mathcal{M}} \omega^{ab}\wedge e^{c}\wedge
e^{d}\varepsilon_{abcd},
\end{equation}
yielding
\begin{equation}\label{EHactionimproved}
\tilde{I}_{EH} = \int_{\mathcal{M}} R^{ab}\wedge e^{c} \wedge e^{d}\varepsilon_{abcd}-
\int_{\mathcal{\partial M}}  \omega^{ab}\wedge e^{c}\wedge e^{d}\varepsilon_{abcd}.
\end{equation}
Now, the variation of $\tilde{I}_{EH}$ reads
\begin{equation}\label{VariationImproved} \delta
\tilde{I}_{EH}|_{\textrm{on shell}} = - 2 \int_{\mathcal{\partial M}} \omega^{ab}\wedge
e^{c}\wedge \delta e^{d}\varepsilon_{abcd},
\end{equation}
confirming that $\tilde{I}_{EH}$ is a proper action principle provided $e^{a}$ is fixed at the
boundary as expected.

The introduction of the boundary term (\ref{omegaAB}) introduces two potential problems. First,
the term is not manifestly invariant under Lorentz transformations since $\omega^{ab}$ transforms
as a connection. Second, it can be divergent at the spatial infinity. To solve both problems one
can add another term,
\begin{equation}\label{Regulator}
\int_{\mathcal{\partial M}}  \omega^{ab}_{0}\wedge e^{c}\wedge e^{d}\varepsilon_{abcd},
\end{equation}
where $\omega^{ab}_{0}$  transforms as a connection in the same fiber as $\omega^{ab}$ but
otherwise satisfying
\[
\delta\omega^{ab}_{0}|_{\mathcal{\partial M}} = 0.
\]
This condition is used since one chooses $\omega^{ab}_{0}$ to represent some particular
background. Thus the role of $\omega^{ab}_{0}$ is to regularize the behavior of the action with
respect to the chosen background. It is direct to prove that $(\omega^{ab}- \omega^{ab}_{0})$
transform as a tensor under Lorentz rotations.

The final expression of $\tilde{I}_{EH}$ is
\begin{equation}\label{Finalexpression}
 \tilde{I}_{EH} = \int_{\mathcal{M}} R^{ab}\wedge e^{c} \wedge e^{d}\varepsilon_{abcd}-
\int_{\mathcal{\partial M}}  (\omega^{ab}- \omega^{ab}_{0})\wedge e^{c}\wedge
e^{d}\varepsilon_{abcd}.
\end{equation}

It is worth to mention that if the vierbein is properly oriented the term added to the action -on
shell- can be rewritten as
\begin{equation}\label{ExtrisicTerm}
\int_{\mathcal{\partial M}} (K-K_{0}) \sqrt{\gamma}\, d^{3}y,
\end{equation}
where $K$ is the trace of the extrinsic curvature of either $\Sigma_{\pm}$ or $\mathbb{R}\times
\partial \Sigma$ respectively, $\gamma$ the determinant of the induced metric and $y$ an adequate coordinate
system. In this view the boundary term can be considered as a generalization of the term proposed
in \cite{Hawking:1996fd}.

To simplify the notation from now on
\[
(\omega^{ab}- \omega^{ab}_{0}) = \hat{\omega}^{ab}.
\]

\subsection{First order gravity in Hamiltonian}

To proceed one needs to define an adequate vierbein and coordinate system. Here it will be used
the line element \cite{Arnowitt:1962hi}
\begin{equation}\label{ADM}
    ds^{2} = -N^{2} dt^{2} + g_{ij}(N^{i} dt + dx^{i})(N^{j} dt + dx^{j}).
\end{equation}
Now, since the coordinates are split in time and spatial, $x^{\mu} = (t,x^{i})$, one can rewrite
$\tilde{I}_{EH}$ as
\begin{equation}\label{EHactionimprovedsplit}
\tilde{I}_{EH} = \int_{\mathcal{M}} (\dot e^{a}_{i} \, \Omega^{\, i\  j}_{a\ bc} \,
 \omega^{bc}_{j}+\omega^{a b}_{t} J_{a b} + e^{a}_{t} P_{a})\,dt\wedge d^3x + B
\end{equation}
where
\begin{eqnarray*}
B&=&\int_{\mathbb{R} \times \partial \Sigma} 2\,e^{a}_{t}\, \Omega_{a\ bc}^{\, i\
j}\,\hat{\omega}^{bc}_{j} \varepsilon_{imn}\,dt\wedge dx^{m}\wedge dx^{n}\\
  J_{ab} &=& 2\, T^{c}_{ij} e^{d}_{k} \varepsilon_{abcd}\varepsilon^{ijk} \\
  P_{d}&=& 2\, R^{ab}_{ij} e^{c}_{k}\varepsilon_{abcd}\varepsilon^{ijk},\\
  \Omega_{a\ bc}^{\, i\  j} &=& 2 \varepsilon_{abcd} \varepsilon^{ijk} e^d_k.\label{Omega}
\end{eqnarray*}
Note that the action (\ref{EHactionimprovedsplit}) has only boundary terms at $\mathbb{R}\times
\partial \Sigma$ but not at the lids $\Sigma_{\pm}$.

Recalling that the vielbein is fixed at the boundary, \textit{i.e.}, $\delta
e^{a}_{\mu}|_{\partial \mathcal{M}}=0$, the variation of the action with respect to $e^{a}_{t}$
and $\omega^{ab}_{t}$ yields the constraint equations,
\[
P_a=0   \textrm{   and   }     J_{ab}=0.
\]
$J_{ab}$ is the generator of Lorentz transformations and $P_{a}$ is the generator of translations
\cite{Banados:1997hs}.

To continue one needs to define the vierbein. Among the different vierbeine that give rise to
Eq.(\ref{ADM}) here, because it significatively simplifies the computations, will be used
\cite{Peldan:1994hi}
\begin{equation}\label{ADMlabel}
    e^{a}_{t} = N \eta^{a} + N^{i} e^{a}_{i}\textrm{  and   }e^{a}_{i} = e^a_{i},
\end{equation}
with
\[
\eta^{a} e_{ai} = 0,\textrm{       }e_{ai}e^{a}_{j}= g_{ij},\textrm{ and    }\eta^{a}\eta_{a}=-1.
\]
$\eta_{a}$ is the unitarian vector normal to the $t=cont.$ slices $\Sigma$. In four dimensions
$\eta^{a}$ can be constructed as
 \[\eta_{a} = \frac{1}{6\sqrt{g}}\,
\varepsilon_{abcd}\,e^{b}_{i}e^{c}_{j}e^{d}_{k}\varepsilon^{ijk},\] where $g = \det{g_{ij}}$.

Using the projection $e^{a}_{t}$ along the $N$ and $N^{i}$ the action can be rewritten as
\begin{equation}\label{Improved}
 \tilde{I}_{EH} = \int_{\mathcal{M}} \dot e^{a}_{i} \, \Omega^{\, i\  j}_{a\ bc} \,
 \omega^{bc}_{j}+ N H_{\perp}+ N^{i} H_{i} + \omega^{ab}_{t} J_{ab} + B,
\end{equation}
where $H_{\perp}$ and $H_{i}$ are the projections of $P_{a}$ along the $\eta^{a}$ and $e^{a}_{i}$
respectively. $N$ and $N^{i}$ are Lagrange multipliers \footnote{One can confirm that the
transformation of fields (\ref{ADMlabel}), \textit{i.e.},
\[ (e^{a}_{t},e^{a}_{i}) \rightarrow
(N,N^{i},e^{a}_{i}),
\]
does not change the measure of the path order integral \cite{Aros:2003bi}.}.

$B$ in Eq.(\ref{Improved}) stands for the boundary term
\[ B = \int_{\mathbb{R} \times \partial \Sigma}  (N \eta^{a} + N^{l} e^{a}_{l})  \left(2\,\Omega_{a\ bc}^{\, i\
j}\,\hat{\omega}^{bc}_{j} \varepsilon_{imn}\right)\,dt\wedge dx^{m}\wedge dx^{n}.
\]

\subsection{Transformation}

To isolate the conjugate momenta of the 12 $e_{i}^{a}$'s, contained in the 18 $\omega _{i}^{ab}$'s
here is introduced the following projection \cite{Contreras:1999je}
\begin{equation}\label{DecomposingTheMomentum}
\omega_{k}^{ab}=\Theta _{k\ \ j}^{ab\ c}\ \pi _{c}^{j}+U_{k}^{ab\ \ mn}\ \lambda _{mn}.
\end{equation}
Note that this also gives rise to others 6 auxiliary fields $\lambda _{mn}$ (and their 6 conjugate
fields $\rho ^{mn}$). $\lambda _{mn}$  (and $\rho ^{mn}$) is symmetric with $m,n =1,2,3$. $\Theta
$ and $U$ are given by
\begin{equation}
\Theta _{i\ \ j}^{ab\ c}=\frac{1}{8\sqrt{g}}\left(e_{i}^{[a}\eta
^{b]}e_{j}^{c}-e_{i}^{[a}e_{j}^{b]}\eta ^{c}-2e_{j}^{[a}\eta ^{b]}e_{i}^{c}\right),
\end{equation}
\begin{equation}
U_{k}^{ab\ \ mn}=\frac{1}{2}\delta _{i}^{(m}\ \epsilon ^{n)kl}\ e_{k}^{a}\ e_{l}^{b},
\end{equation}
where the square brackets indicate antisymmetrization.

In addition one can introduce
\begin{equation}
V_{ab\ \ mn}^{k}=\frac{1}{g}\ E_{a}^{r}\ E_{b}^{s}\epsilon _{rs(m}\ \delta _{n)}^{i},
\end{equation}
such that unveiling the relations
\begin{eqnarray}
\Omega _{ab\ c}^{k\ \ i}\ \Theta _{k\ \ j}^{ab\ d}=\delta _{d}^{c}\delta _{j}^{i},\nonumber   \\
\Omega _{ab\ c}^{k\ \ j}\ U_{k}^{ab\ \ mn}=0, \label{orto4} \\
\Theta _{k\ \ j}^{ab\ c}\ V_{ab\ \ mn}^{k}=0,\nonumber  \\
U_{k}^{ab\ \ mn}\ V_{ab\ \ pq}^{k}=\delta _{(pq)}^{(mn)}\nonumber .
\end{eqnarray}

This allows to think of $\Theta $ and $\Omega $ as a collection of twelve vectors - labeled by the
indices $(_{i}^{a})$ and $(_{a}^{i})$ respectively-, in an 18-dimensional vector space with
components $(_{j}^{ab})$, and $(_{ab}^{j})$, respectively. Analogously, $U$ and $V$ correspond to
other six vectors. In this way the orthonormal relations (\ref{orto4}) become the completeness
relation
\begin{equation}
\Theta^{ab \ e}_{i \ \ l} \ \Omega_{cd \ e}^{j \ \ l} + U^{ab \ \ mn}_{i} \ V_{cd \ \ mn}^{j} =
\delta^{[a b]}_{[c d]} \delta^i_j ,  \label{completeness}
\end{equation}
or equivalently
\[
\left(\Theta \, U\right) \left(
\begin{array}{c}
\Omega\\
V
\end{array}\right)
= Id_{18\times 18}
\]
in the 18 dimensional space.

Analogously at the boundary one can define
\[
\hat{\omega} _{k}^{ab}=\Theta _{k\ \ j}^{ab\ c}\ \hat{\pi} _{c}^{j}+U_{k}^{ab\ \ mn}\
\lambda_{mn}.
\]

\subsection{A Hamiltonian expression} Using the decomposition in Eq.(\ref{DecomposingTheMomentum})
the action reads
\begin{eqnarray}
\tilde{I}_{EH}  &=& \int_{\mathcal{M}} (\dot e^{a}_{i} \pi_{a}^{i}+ N H_{\perp}+ N^{i} H_{i} +
\omega^{ab}_{t} J_{ab})\, dt\wedge d^{3}x \nonumber\\
   &+ & B.\label{ImprovedDecomposed}
\end{eqnarray}
Here $\pi_{a}^{i}$ is indeed the conjugate momentum of $e^{a}_{i}$. Furthermore
Eq.(\ref{ImprovedDecomposed}) is a genuine Hamiltonian action principle provided $\delta
e^{a}|_{\partial M}=0$. $H_{\perp}$, $H_{i}$ and $J_{ab}$ are first class constraints
\cite{Banados:1997hs}. For their expressions in terms of the fields see appendix
\ref{ExplicitExpressionsofConstr}.

The Hamiltonian of this theory reads
\begin{equation}\label{HamiltonianFinalExpression}
H=-\int_{\Sigma} N H_{\perp}+ N^{i} H_{i} + \omega^{ab}_{t} J_{ab}\,d^{3}x - \hat{B},
\end{equation}
where
\begin{equation}\label{Boundaryterm}
 \hat{B} = \int_{\partial \Sigma} \left( N \eta^{a} \hat{\pi}^{i}_{a} + N^{l} e^{a}_{l}
\hat{\pi}^{i}_{a}\right)\varepsilon_{imn}
    dx^{m}\wedge dx^{n}.
\end{equation}

Recalling that the constraints vanishes on shell then the Hamiltonian
(\ref{HamiltonianFinalExpression}) becomes merely the boundary term, $H_{on\, shell}= -\hat{B}$.
This last observation will be essential to develop an expression for the energy in the next
sections.

\subsection{Geometry and coordinates at the boundary}

The boundary $\mathbb{R}\times \partial\Sigma$ has a metric of the form
\begin{equation}\label{ADMBoundary}
    ds^{2} = -N^{2} dt^{2} + h_{mn}(V^{m} dt + d\sigma^{m})(V^{n} dt + d\sigma^{n}),
\end{equation}
where $\sigma^{m}$, with $m=2,3$, are the coordinates of slice at $t=const.$ of this boundary.
Since the boundary can be described as a surface $x^{\mu}(t,\sigma^{m})$ one can define a set of
(co-)vectors which give rise to metric (\ref{ADMBoundary}). This set reads
\begin{eqnarray}
  e^{a}_{t} &=& N \eta^{a} + V^{m} e^{a}_{m} \nonumber\\
  e^{a}_{m} &=& e^a_{m}, \label{ADMBoundarylabel}
\end{eqnarray}
where the projections are made by
\[ V^{m} = N^{i} \left.\frac{\partial \sigma^{m}}{\partial x^{i}}\right|_{\mathbb{R}\times
\partial\Sigma} \textrm{ and } e^{a}_{m} =  e^{a}_{i}\left.\frac{\partial x^{i}}{\partial
\sigma^{m}}\right|_{\mathbb{R}\times \partial\Sigma}
\]

To complete this analysis usually is introduced the unitarian vector $n^{a}$ which is normal to
the boundary $\mathbb{R}\times \partial \Sigma$, \textit{i.e.},
\[
n_{a}\eta^{a}=0 \textrm{ , } n_{a} e^{a}_{m} = 0 \textrm{ and } n^{a}n_{a}=1.
\]
This vector can written as
\begin{equation}\label{VectorN}
n_{a} = \frac{1}{2 \sqrt{\gamma}}\varepsilon_{abcd}\, \eta^{b} e^{c}_{m}
e^{d}_{n}\varepsilon^{mn},
\end{equation}
where $\gamma=N^{2}h$ is the determinant of the induced metric (\ref{ADMBoundary}) on
$\mathbb{R}\times\partial\Sigma$. Note that $n^{a}$ is only a functional of $\eta_{a}$ and
$e^{a}_{m}$. Note that also one can obtain
\begin{equation}\label{VectorEtaB}
\eta_{a} = \frac{1}{2 \sqrt{h}}\varepsilon_{abcd}\, n^{b} e^{c}_{m} e^{d}_{n}\varepsilon^{mn},
\end{equation}

\subsection{Energy and Momentum} Using the projections (\ref{ADMBoundarylabel}) the boundary term
reads
\begin{equation}\label{projectedBoundaryTerm}
B = \int_{\mathbb{R} \times \partial \Sigma} \left( N \eta^{a}  + V^{m} e^{a}_{m}
\right)(\hat{\pi}_{a}\cdot n)dt \wedge d^{2}\sigma.
\end{equation}
where $(\hat{\pi}_{a}\cdot n)$ represents the projection $\hat{\pi}_{a}^{i}n_{i}$ at the boundary.

Following the generalization of the Hamilton Jacobi equations proposed in \cite{Brown:1992br} one
can define an expression for the energy. Since the fields at the boundary are
$(N,V^{m},e^{a}_{m})$ here it is advisable to directly variate the action with respect each of
dynamical fields
\begin{eqnarray}
\left.\frac{\delta \tilde{I}_{EH}}{\delta N}\right|_{on shell}=\eta^{a}(\hat{\pi}_{a}\cdot n), & &
\left.\frac{\delta \tilde{I}_{EH}}{\delta V^{m}}\right|_{on shell} =
  e^{a}_{m} (\hat{\pi}_{a}\cdot n)\nonumber,\\
\left.\frac{\delta \tilde{I}_{EH}}{\delta e^{a}_{m}}\right|_{on shell} = \tau_{a}^{m}.&&
\label{variationAttheBoundary}
\end{eqnarray}
Note that $\tau^{m}_{a}$ is not a squared matrix.

A definition of energy can be obtained by integrating Eq.(\ref{variationAttheBoundary})
\begin{equation}\label{Energy}
    E = -\int_{\partial \Sigma} \eta^{a}(\hat{\pi}_{a}\cdot n)d^{2}\sigma.
\end{equation}
It is straightforward to show that this expression indeed recovers the mass of Schwarzschild or
Reissner N{\o}rdstrom solutions provided $\omega^{ab}_{0}$ correspond to Minkowski space.

Likewise one can define the \textit{momentum}
\begin{equation}\label{Momentum}
    P_{m} = \int_{\partial \Sigma} e^{a}_{m} (\hat{\pi}_{a}\cdot n) d^{2}\sigma,
\end{equation}
and an intrinsic \textit{energy momentum} tensor
\begin{equation}\label{IntrisicEnergyMomentumTensor}
T^{m}_{a} = \int_{\partial \Sigma}d^{2}\sigma \tau_{a}^{m}.
\end{equation}

One can define energy density $e=-\eta^{a}(\hat{\pi}_{a}\cdot n)$ and the momentum density
$p_{m}=e^{a}_{m} (\hat{\pi}_{a}\cdot n) $.

Note that with these definitions the Hamiltonian can be written as
\begin{equation}\label{HamiltonianInterpreted}
H = H_{\textrm{bulk}} + \int_{\partial \Sigma} ( e N - V^{m}p_{m})\, d^{2}\sigma.
\end{equation}
It is interesting to compare this result with the analogous in \cite{Brown:1992br}, since the
underlying content of fields is different. For instance after a straightforward computation the
expression of the energy can be split as
\begin{equation}\label{Edecomposition}
\left.\eta^{a}(\hat{\pi}_{a}\cdot n)\right|_{\partial \Sigma}d^{2}\sigma =
[(\mathbf{k}-\mathbf{k}_{0}) + n_{a} K^{a b}_{\hspace{2ex}c} h^{m}_{b} h_{m c}]
\sqrt{h}d^{2}\sigma,
\end{equation}
where $\mathbf{k}$ is the trace of the intrinsic curvature of $\partial \Sigma$ immersed in
$\Sigma$ and $h^{m}_{b}$ is the projector from $\mathcal{M}$ into $\partial\Sigma$. The first part
Eq.(\ref{Edecomposition}) recovers the expression for the energy in \cite{Brown:1992br}. However
the second term is intrinsical to first order gravity, since it explicitly depends on the
contorsion tensor $K^{a b}_{\hspace{2ex}c}$. In absence of fermions, since $K^{ab}_{c}$ vanishes
on shell, the expression in \cite{Brown:1992br} is formally recovered as expected.

For the momentum $P_{m}$ the expression can be written as
\begin{equation}\label{Pdecomp}
    P_{m} = -2\int_{\partial \Sigma} h_{m i} \hat{\pi}^{ij} n_{j} d^{2}\sigma + F(K^{a b}_{\hspace{2ex}c},h^{a}_{m})
\end{equation}
where $ h_{m i}$ is the project from $\Sigma$ into $\partial \Sigma$, and $\pi^{ij}$ is the
metrical expression for the momentum. $F(K^{a b}_{\hspace{2ex}c},h^{a}_{m})$ is lineal function of
the contorsion tensor, which vanishes for $K^{ab}_{c}=0$. Therefore the first part of
Eq.(\ref{Pdecomp}) actually recovers the metrical expression and the rest depends on the
contorsion tensor, thus again in absence of fermions the expression in \cite{Brown:1992br} is
formally recovered. Finally one can show that the projection of $T^{a}_{m}$ along $e^{a}_{i}$
matches the metrical expressions in \cite{Brown:1992br} provided $K^{ab}_{c}=0$.

\subsection{Canonical ensemble action} The variation of the action (\ref{ImprovedDecomposed}) can
be cumbersome in terms of the phase space fields, however recognizing that on shell the variation
is merely given by Eq.(\ref{VariationImproved}) one obtains that
\begin{eqnarray}
\delta \tilde{I}_{EH}  &=& \int_{\Sigma_{\pm}} \hat{\pi}^{i}_{a} \delta (e^{a}_{i}) \label{FixofTheEnsemble}\\
 &+& \int_{\mathbb{R}\times
\partial \Sigma} (e\delta N -  p_{m} \delta V^{m}  + \tau^{m}_{a} \delta e^{a}_{m})\,dt\wedge
d^{2}\sigma.\nonumber
\end{eqnarray}
The first term basically represents the generalization of the standard $p\,\delta
x|_{t_{i}}^{t_{f}}$ in any $0+1$ Lagrangian, in this case in the lids $\Sigma_{\pm}$.

The next term in Eq.(\ref{FixofTheEnsemble}) shows that in the variational principle
(\ref{EHactionimprovedsplit}) the energy, as define in Eq.(\ref{Energy}), is not fixed, but the
lapse $N$. The fixing of $N$ in turns fixes the scales of time, and thus the period in the
Euclidean version of the $\mathcal{M}$, \textit{i.e.},
\begin{equation}\label{FixingofPeriod}
    \beta = i \oint_{\partial \Sigma} dt N.
\end{equation}

Note that when $\partial \Sigma$ is composed by more than a single surface, like in a black hole
geometry, then one can fix $N=N_{0}$ at only one of those boundaries. In this work $N$ will be
fixed at infinity $\partial \Sigma_{\infty}$ and although it is not formally necessary as $N
|_{\partial \Sigma_{\infty}}= N_{\infty}=1$.

The combination of an unconstrained energy in Eq.(\ref{FixofTheEnsemble}) and the fixing of
$\beta$ suggests that the action (\ref{ImprovedDecomposed}) might be suitable for the
(grand)-canonical ensemble. To confirm this statement one can study the statistical mechanics
framework around the charges (Eqs.(\ref{Energy},\ref{Momentum})).

To proceed is necessary to consider a particular solution. Here the most general stationary black
hole in vacuum (with $\Lambda=0$) will be considered, the Kerr solution. Since this solution has
mass $M$ and angular momentum $J$ one must note that this solution is suitable for the
\textit{grand canonical ensemble}. As background configuration it has been chosen the Minkowski
space.

At zero order approximation on the path order integral arises the relation for the partition
function in the grand canonical ensemble
\begin{equation}\label{PartitionFunction}
\ln(\mathbf{Z})= \beta \bar{E} + \beta\Omega \bar{J} - S(\beta,\Omega) \approx
\mathbf{I}^{E}_{\beta,\Omega}|_{\textrm{on shell}} + O(x^{2}),
\end{equation}
where $\Omega$ in this case corresponds to the value of the angular velocity of the horizon.

The connection between the statistical mechanics and the boundary terms becomes clearer once one
notes that any stationary solution satisfies $\dot{e}^{a}_{i}=0$, therefore the action merely
reduces to the boundary terms
\begin{equation}\label{Equality}
\mathbf{I}_{E}|_{\textrm{on shell}}= B|_{\mathbb{R}\times \partial \Sigma}.
\end{equation}

As mention before in this case the horizon must be considered as an internal boundary,
\textit{i.e.}, $\partial \Sigma=\partial \Sigma_{\infty} \oplus
\partial \Sigma_{H}$. Therefore,
\begin{equation}\label{CanonicalEnsemble}
\mathbf{I}_{E}|_{\textrm{on shell}} = B|_{ \mathbb{R}\times \partial \Sigma_{\infty} }- B|_{
\mathbb{R}\times \partial \Sigma_{H}}.
\end{equation}

From the definitions in the previous sections one obtains that expression at infinity give rise to
the value of the charges,
\[
B|_{x\rightarrow \mathbb{R}\times \partial \Sigma_{\infty} } = \beta (\bar{E} + \Omega \bar{J}).
\]

If the action (\ref{Finalexpression}) is truly sensible for the canonical ensemble then, by
connecting Eq.(\ref{PartitionFunction}) and Eq.(\ref{Equality}), the entropy must be given by
\[
S = B|_{x\rightarrow \mathbb{R}\times \partial \Sigma_{H}}.
\]

To compute the value of $S$ one needs to define some general properties of the horizon first. Near
the horizon the Euclidean metric becomes \cite{Gibbons:1979xm}
\begin{equation}\label{NearHorizonMetric}
    ds^{2}|_{x\rightarrow \mathbb{R}\times \partial \Sigma} \approx N^{2} d\tau^2 + h_{ij} dx^{i} dx^{j},
\end{equation}
which in terms of the vierbein reads
\[ (N^{i} e^{a}_{i})|_{x\rightarrow \mathbb{R}\times \partial \Sigma_{H}} \approx 0.
\]

This general consideration permits to confirm, after computing the corresponding asymptotical
limit at the horizon of $\omega^{ab}$, that the standard area law
\begin{equation}\label{AnExpEntropy}
    S = \lim_{x\rightarrow \mathbb{R}\times \partial \Sigma_{H}} \int \hat{\omega}^{ab}\wedge e^{c}\wedge
e^{d}\varepsilon_{abcd} \approx \frac{A}{4}
\end{equation}
is recovered. This confirms also that the principle action proposed in Eq.(\ref{Finalexpression})
indeed corresponds to the (grand) canonical action.

\section{First order boundary terms with $\Lambda <0$}

The boundary conditions in spaces with a negative cosmological constant within first order gravity
has been observed  to be fundamentally different \cite{Aros:2001gz}. In this case is more adequate
to impose boundary conditions  $\omega^{ab}$ and its derivative than on the vierbein as in
$\Lambda=0$. This leads to proceed in a generalization of $\omega$-frame to study this case.

\subsection{Einstein Hilbert action with $\Lambda<0$}

To initiate the discussion one can consider the four dimensional case with a negative cosmological
constant. The four dimensional Einstein Hilbert action with a negative cosmological constant in
first order formalism reads
\begin{equation}\label{EHactionnegative}
\mathbf{I}_{EH} = \frac{1}{64G} \int_{\mathcal{M}}( 2R^{ab}\wedge e^{c}\wedge e^{d} + l^{-2}
e^{a}\wedge e^{b}\wedge e^{c} \wedge e^{d})\varepsilon_{abcd},
\end{equation}
where the cosmological constant has been written in terms of the AdS radius as $\Lambda = - 1/
(3l^2)$.

The variation of Eq.(\ref{EHactionnegative}) yields
\begin{equation}\label{EinsteinEquationsNegative}
\delta e^{d} \rightarrow \left(R^{ab}\wedge e^{c}+ \frac{1}{l^2} e^{a}\wedge e^{b} \wedge
e^{c}\right)\varepsilon_{abcd}=0,
\end{equation}
and the equation $T^{a}=0$ already obtained in Eq. (\ref{TorsionEquations}). When $T^{a}$ is
replaced on Eq.(\ref{EinsteinEquationsNegative}) it becomes the standard Einstein equations with a
negative cosmological constant.

The presence of a negative cosmological constant gives rise to several technicalities, in
particular the usual expressions of the charges, as for instance the Komar's potentials
\cite{Komar1959}, become divergent. This problem has been addressed in many works (for instance
\cite{Hawking:1996fd,Henneaux:1999ct}) and is particular important in the context of the AdS/CFT
conjecture (See for instance \cite{Emparan:1999pm, Skenderis:2002wp}).

In \cite{Aros:1999id,Aros:2001gz,Aros:2002ub} was discussed a set of boundary conditions that
allows to transform Eq.(\ref{EHaction}) into a proper action principle. Under this boundary
conditions is added to Eq.(\ref{EHactionnegative}) the term
\begin{equation}\label{EulerTerm}
\mathbf{E}=\frac{l^2}{64G}\int_{\mathcal{M}} R^{ab}  R^{cd}\varepsilon_{abcd},
\end{equation}
whose variation is a total derivative and thus it does not alter the equations of motion.
Eq.(\ref{EulerTerm}) is usually called the Euler term, but it is not the Euler number of the
manifold though.

The new action principle reads
\begin{equation}\label{RR1}
\mathbf{I}_{EH} = \frac{l^2}{64G} \int_{\mathcal{M}} \bar{R}^{ab} \bar{R}^{cd}\varepsilon_{abcd}
\end{equation}
with $\bar{R}^{ab}= R^{ab}+l^{-2} e^{a} e^{b}$. On shell the variation of Eq.(\ref{RR1}) yields
\begin{equation}\label{BoundarytermNegative}
\delta \mathbf{I}_{EH} = \frac{l^2}{32G}\int_{\partial \mathcal{M}} \delta
\omega^{ab}\bar{R}^{cd}\varepsilon_{abcd}.
\end{equation}

The addition of Eq.(\ref{EulerTerm}) is made to obtain an action principle suitable for any
asymptotically locally anti de Sitter (\textbf{ALAdS}) space. To confirm that one can note that
generically $\bar{R}^{ab}(x)|_{x\rightarrow\partial \Sigma_{\infty}} \rightarrow 0$ for any ALAdS
space, and thus Eq.(\ref{Boundaryterm}) has no contributions from the asymptotical spatial region
$\mathbb{R}\times \partial \Sigma_{\infty}$. In the other boundaries of $\mathcal{M}$ an adequate
boundary condition is to fix the spin connection.

One of the surfaces in which the spin connection is to be fixed is the horizon. However the
horizon requires some special attention, since to fix the spin connection at the horizon is
connected with the fixing of the temperature of the black hole \cite{Aros:2001gz}. To see that one
first must recall that the horizon of a stationary black hole is the surface (in $\mathcal{M}$)
where $\xi=\xi^{\mu}\partial_{\mu}$ the horizon generator, a time like Killing vector, becomes
light like. Next the temperature of the black hole can be read from the relation
\begin{equation}\label{Temperature}
I_{\xi} \omega^{a}_{\hspace{1ex} b} \xi^{b}|_{\mathbb{R}\times
\partial\Sigma_{H}} = \kappa \xi^{b},
\end{equation}
where $\kappa$ is the surface gravity. The temperature is given by $T=\kappa/4\pi$. In this way
the fixing of the spin connection at the horizon determines the temperature. The relation
(\ref{Temperature}) is the first order version of the relation
\[
\xi^{\mu}\nabla_{\mu}(\xi^{\nu})|_{\mathbb{R}\times
\partial\Sigma_{H}} = \kappa \xi^{\nu}
\]
obtained in \cite{Wald:1993nt}. By a simple translation between metric and vielbein formalisms
\cite{yvonne} one can prove that the fixing of the spin connection also fixes the extrinsic
curvature.

Note that here it was not necessary to explicitly require the smoothness of the Euclidean manifold
at the horizon to obtain the temperature.

\subsection{The Hamiltonian}
The introduction of the coordinate system described in Eq.(\ref{ADM}) leads to rewrite the
Lagrangian in Eq.(\ref{RR1}) as
\begin{eqnarray}
\bar{R}^{ab}\bar{R}^{cd}\varepsilon_{abcd} &=& (2
\dot{\omega}^{ab}_{i}\bar{R}^{cd}_{jk}\varepsilon_{abcd}\epsilon^{ijk}\nonumber\\
&+& \omega^{ab}_{t} J_{ab} + e^{d}_{t} P_{d} + \partial_{i} J^{i}) d^4x\label{RRrewritten}
\end{eqnarray}
with
\begin{eqnarray*}
  J_{ab} &=& 4\, T^{c}_{ij} e^{d}_{k} \varepsilon_{abcd}\epsilon^{ijk} \\
  P_{d}&=& 4\, \bar{R}^{ab}_{ij} e^{c}_{k}\varepsilon_{abcd}\epsilon^{ijk} \\
   J^{i}&=& 2\, \omega^{ab}_{t}\bar{R}^{cd}_{jk}\varepsilon_{abcd}\epsilon^{ijk}
\end{eqnarray*}

Remarkably the Lagrangian has only a boundary term at $\mathbb{R}\times\partial \Sigma$ but not at
the lids of the cylinder.

The introduction of the vierbein (\ref{ADMlabel}) yields
\begin{eqnarray}
\mathbf{I}_{EH} &=& \frac{l^2}{64G}\int \left( 2\dot{\omega}^{ab}_{i}P_{ab}^{i}  +\omega^{ab}_{t} J_{ab}\right.\nonumber \\
&+& \left.N H_{\perp} + N^{i} H_{i} + \partial_{i} J^{i}\right) d^4x\label{RRonNN},
\end{eqnarray}
where
\begin{eqnarray*}
P_{ab}^{i}&=&\bar{R}^{cd}_{jk}\varepsilon_{abcd}\epsilon^{ijk}\\
H_{\perp}&=&  P^{k}_{cd}e^{c}_{k}\eta^{d}= 6\sqrt{g}\bar{R}^{cd}_{ij} E^{i}_{c}E^{j}_{d}\\
H_{l} &=& P^{k}_{cd}e^{c}_{k}e^{d}_{l} = \varepsilon_{abcd}\left(\bar{R}^{ab}_{ij}
e^{c}_{k}\epsilon^{ijk}\right) e^{d}_{l},
\end{eqnarray*}
with $E^{\mu}_{a}$ the inverse of the vierbein.

Eq.(\ref{RRonNN}) defines $P_{ab}^{i}$ as the conjugate the momentum of $\omega^{ab}_{i}$,
however, the expression of the Hamiltonian is incomplete.

For $\Lambda <0 $, given that the boundary conditions depend on the spin connection, the
expressions have been studied in an extension of the $\omega$ frame. In the original $\omega$
frame (\textit{i.e.} $\Lambda=0$ \cite{Contreras:1999je}) second class constraints arise because
$P_{ab}^{i}$, which has 18 component, formally depends on $e^{a}_{i}$ with only 12 component. Here
with $\Lambda<0$, and for the same reasons, second order constraints arise as well. Thus besides
the terms in Eq.(\ref{RRonNN}) is necessary to added these second class constraints to complete
the Hamiltonian \cite{ArosInProgress}. Finally the Hamiltonian reads
\begin{eqnarray}
\mathbf{H} &=& \frac{l^2}{64G}\int_{\Sigma} \left(\omega^{ab}_{t} J_{ab} + N H_{\perp} + N^{i}
H_{i} +
\Phi^{ij}\mu_{ij}\right) \,d^3 x\nonumber\\
   &+& \frac{l^2}{32G} \int_{\partial \Sigma} \omega^{ab}_{t}\bar{R}^{cd}_{jk}\varepsilon_{abcd}
     dx^{j}\wedge dx^{j}\label{Hamiltonian}
\end{eqnarray}
where $\Phi^{ij}$ are 6 second order constraints, whose expression reads
\begin{equation}\label{Phiij}
\Phi^{ii'} = \varepsilon^{aba'b'}\hat{P}^{i}_{ab}\hat{P}^{i'}_{a'b'}
\end{equation}
with
\[
\hat{P}^{i}_{ab} = P^{i}_{ab}-\epsilon^{ijk}\varepsilon_{abcd}R^{cd}_{jk} \simeq (E_{a}^{t}
E_{b}^{i}-E_{b}^{t} E_{a}^{i})|e|.
\]

This definition of momentum allows to rewrite the generator of Lorentz rotations as
\begin{equation}\label{Jab}
    J_{ab} = D_{i}(P_{ab}^{i}) = D_{i}(\hat{P}_{ab}^{i}).
\end{equation}
The expressions of $H_{i}$ and $H_{\perp}$ in terms of the fields are cumbersome, not very
illustrative, and furthermore irrelevant for what follows.

\subsection{Noether Charges} The vanishing of the constraints on shell implies that
\begin{equation}\label{HamiltonianOnShell}
\mathbf{H}|_{\textrm{on shell}} =\frac{l^2}{32G} \int_{\partial \Sigma_{\infty}-\partial
\Sigma_{H}} \omega^{ab}_{t}\bar{R}^{cd}_{jk}\varepsilon_{abcd}\, dx^{j}\wedge dx^{j}.
\end{equation}
This is the expression that is necessary to connect with the value of mass, entropy and angular
momentum. This connection will be done through the value of Noether charges associated with the
Killing vectors of the solution. Although the interpretation of mass or angular momentum as
Noether charges was one of the original ideas of E. Noether, it is worth to stress that their
connection with entropy was proposed in \cite{Wald:1993nt}.

In \cite{Aros:1999id} was found that the expression for the Noether charge associated with any
Killing vector $\eta$ reads
\begin{equation}\label{NoetherCharge}
    Q_{\eta} = \frac{l^2}{32G}\int_{\partial \Sigma} I_{\eta}\omega^{ab} \bar{R}^{cd}
    \varepsilon_{abcd}.
\end{equation}
where $I_{\xi}$ stands for the projector along the vector $\eta$, in this case
$I_{\eta}\omega^{ab}=\eta^{\mu}\omega^{ab}_{\hspace{2ex}\mu}$.  The mass or the angular momentum
of the solution can be obtained from Eq.(\ref{NoetherCharge}) as the asymptotical value at
$\partial \Sigma_{\infty}$ and provided $\eta$ be the Killing vector associated with time or
rotational symmetries respectively. The evaluation at the horizon for $\xi$, the horizon
generator, leads to the entropy.

\subsection{An expression for the Hamiltonian} To study this ideas one can choose a particular but
representative solution. The most general stationary black hole in vacuum with a negative
cosmological constant will be analyzed, the Kerr-AdS solution. Since this solution has angular
momentum the canonical ensemble is replaced by the grand canonical ensemble.

The most general Killing vector of this solution reads
\[
\eta = \alpha \xi_{t} + \beta \xi_{\phi}
\]
where $\alpha$ and $\beta$ are constant and $\xi_{t}$ and $\xi_{\phi}$ are the Killing vectors
associated with the time and rotational invariance respectively. In terms of these symmetries the
Killing vector that defines the horizon, $\xi$, reads
\begin{equation}\label{KillingVectorHorizon}
\xi = \xi_{t} + \Omega \xi_{\phi},
\end{equation}
where $\Omega$ is the \textit{angular velocity} of the horizon \cite{Hawking:1998kw}.

Usually the Kerr-AdS solution is presented in the Boyer-Lindquist (\textbf{BL}) coordinates (See
appendix \ref{KerrAdS}). The projection of the spin connection in these coordinates reads
\begin{equation}\label{omegaT}
    I_{\xi} \omega^{ab} = \omega^{ab}_{t} + \Omega \,\omega^{ab}_{\phi}.
\end{equation}
Now, by noting that the expression of the Hamiltonian in Eq.(\ref{HamiltonianOnShell}) formally
equals the expression of the Noether charge for $\xi_{t}=\partial_{t}$ in the BL coordinates one
obtains
\begin{equation}\label{HamiltonianOnNoetherChargesI}
\mathbf{H}|_{\textrm{on shell}} = M - \frac{l^2}{32G}\int_{\partial \Sigma_{H}}
\left(I_{\xi}\omega^{ab} - \Omega \,\omega^{ab}_{\phi}\right) \bar{R}^{cd}\varepsilon_{abcd}.
\end{equation}
Finally, after a long but straightforward computation, one can prove that the last term is the
angular momentum of the solution. Thus,
\begin{equation}\label{HamiltonianOnNoetherChargesIII}
\mathbf{H}|_{\textrm{on shell}} = M + \Omega J - Q_{\xi}|_{\Sigma_{H}}.
\end{equation}

Using zero order approximation on the generalization of Eq.(\ref{PartitionFunctionFeynman}) to the
grand canonical ensemble one obtains,
\[ \ln(\mathbf{Z})= \beta \bar{E} +
\beta\Omega \bar{J} - S(\beta,\Omega) \approx -\mathbf{I}^{E}_{\beta,\Omega}|_{\textrm{on shell}}
+ O(x^{2}).
\]

Since in this case the solution in discussion is a stationary black hole then
$\dot{\omega}^{ab}_{i}=0$ which in turns implies that
\begin{equation}\label{EqualityNegative}
-\mathbf{I}_{E}|_{\textrm{on shell}} = -\beta \mathbf{H}|_{\textrm{on shell}} \approx
\ln(\mathbf{Z}).
\end{equation}
Note that the right hand side also can be obtained as the zero order approximation of
Eq.(\ref{PartitionFunction}).

Combining these results is possible to recognize that the entropy is
\[
S-S_{0} = \beta \left(\frac{l^2}{32G}\int_{\partial \Sigma_{H}}I_{\xi} \omega^{ab} \bar{R}^{cd}
\varepsilon_{abcd}\right).
\]
where $S_{0}$ stands for possible higher order corrections to the value of entropy but can not
depend on the values of the extensive variables. This result is equivalent to the one obtained in
\cite{Aros:2001gz}. Computing explicitly $S$ yields to the usual
\[
S = \frac{1}{4G} A
\]
where $A$ stands for the area of the horizon.

\subsection{Higher Dimensions}

In higher dimensions besides the EH theory of gravities there are several other sensible
gravitational theories, with second order equation of motion for the metric. Among them one
important group are the usually called Lovelock gravities \cite{Lovelock:1971yv}. First order
formalism is particular suitable for their study, since here the first order nature of the
equation of motion for ($e^{a},\omega^{ab}$) is manifest, and thus the second order nature of the
metric ones. In absence of fermions in general these gravities have only solutions with vanishing
torsion\footnote{ There is however the so called Chern Simons gravities where, even in absence of
fermions, the most general solution, in first order formalism, has  $T^{a}\neq 0$.}.

Schematically the Lagrangian of Lovelock gravities reads
\begin{equation}\label{LL}
\mathbf{L} = \sum_{p=0}^{[\frac{d-1}{2}]} \frac{\alpha_{p}}{l^{d-2p}} R^{a_{1} a_{2}} \ldots
R^{a_{2k-1} a_{2p}} e^{a_{2p+1}} \ldots e^{a_{d}}\varepsilon_{a_{1}\ldots a_{d}}
\end{equation}
where $\alpha_{p}$ are arbitrary constants and $[\,]$ stands for function \textit{integer part of
the argument}.

In general this Lagrangian has solutions with multiple cosmological constants. This can be
considered a problem since that produces unstable geometries that can tunnel between the different
cosmological constants. To solve this one can restrict the $\alpha_{p}$ coefficients such that
only a single cosmological constant exist.

With a single cosmological constant the form of the equation of motion associated with $\delta
e^{a}$ is
\begin{equation}\label{LEH}
\bar{R}^{a_{1} a_{2}} \ldots \bar{R}^{a_{2k-1} a_{2k}} e^{a_{2k+1}} \ldots
e^{a_{d-1}}\varepsilon_{a_{1}\ldots a_{d}} = 0
\end{equation}
which can be obtained provided
\begin{equation}\label{alphaCoeffients}
\alpha_{p}= \left\{ \begin{array}{ll}
                   \frac{1}{d-2p} \left(\begin{array}{c}
                               k \\
                               p \end{array} \right)  & p\leq k \\
                          0 & p>k \\
                        \end{array}\right..
\end{equation}

One can prove in general that the solutions of these restricted Lovelock theories are ALAdS
spaces.

Remarkably the results for the four dimensional EH gravity in the previous sections can be easily
extended to these restricted Lovelock gravities in even dimensions. In odd dimensions, however,
this is not direct because there is no generalization of Eq.(\ref{EulerTerm}), namely there is no
Euler term.

In general in $d=2n$ dimensions for an ALAdS space one can use the same boundary conditions
already proposed. To do that one adds to the now restricted $\mathbf{L}$ the term
\begin{equation}\label{ELL}
\mathbf{E}_{2n} = \frac{{\mathcal{K}}}{l^{2n}} \int_{\mathcal{M}} R^{a_{1}a_{2}}\ldots
R^{a_{2n-1}a_{2n}} \varepsilon_{a_{1}\ldots a_{2n}}
\end{equation}
where
\[
{\mathcal{K}} = -\sum_{p=0}^{n-1} (-1)^{p}\alpha_{p} = -\frac{1}{2n \left(\begin{array}{c}
                               k-n  \\
                               n\end{array} \right)}.
\]

As outcome of this addition the variation of the new action principle -on shell- yields a boundary
term that at the asymptotically spatial region ($\mathbb{R}\times \partial \Sigma_{\infty}$) of
any ALAdS space, behaves as
\begin{equation}\label{LLBoundaryTerm}
    \Theta_{x\rightarrow \partial \Sigma_{\infty}} \sim \delta\omega\,\bar{R}^{k-1} e^{2n-2k}\rightarrow 0,
\end{equation}
and thus the asymptotical spatial region $\mathbb{R}\times \partial \Sigma_{\infty}$ does not
contribute to the variation of the action. A proper boundary condition at the other boundaries is
$\delta \omega^{ab}=0$. Analogous to the four dimensional case if a black hole geometry is
considered the temperature is fixed by this boundary condition.

Following analogous steps as the four dimensional case one obtains that on-shell the Hamiltonian
is merely the boundary term
\begin{eqnarray}
 \mathbf{H}_{On shell} &=& \int_{\partial \Sigma} \sum_{p=1}^{n-1}  \frac{p\,\alpha_{p}}{l^{2n-2p}} \omega_{t} R^{p-1}
e^{2n-2p} \nonumber \\
   &+& n{\mathcal{K}} \int_{\partial \Sigma} \omega_{t} R^{n-1}.\label{LLHamiltonian}
\end{eqnarray}

To proceed one needs to consider a particular solution. Here it will be considered the topological
black holes studied in Ref.\cite{Aros:2000ij}. They are described in Appendix
\ref{TopologicalBHSec}.

Recalling the expression of the Noether charge which generically reads
\begin{equation}\label{LLNoetherCharge}
    Q_{\xi} = \int_{\partial \Sigma} I_{\xi} \omega^{ab} \frac{\partial\mathbf{L}}{\partial
    R^{ab}},
\end{equation}
where
\[
\frac{\partial\mathbf{L}}{\partial R} = \sum_{p=1}^{n-1} \frac{ p\,\alpha_{p}}{l^{2n-2p}} R^{p-1}
e^{2n-2p} + n{\mathcal{K}} R^{n-1},\
\]
one can connect the Lagrangian and Hamiltonian results. The horizon generator $\xi$ of these
solution is merely $\xi=\partial_{t}$ (See appendix \ref{TopologicalBHSec}), thus it is direct to
identify the part of the Hamiltonian (\ref{LLHamiltonian}) at $\partial \Sigma_{\infty}$ as the
Noether charge associated with the time symmetry, and so with the mass of the solution.

Since these are static solutions is also satisfied $\dot \omega^{ab}_{i}=0$ and so
\[-\mathbf{I}_{E}|_{\textrm{on shell}} = -\beta \mathbf{H}|_{\textrm{on shell}}.\]

After analogous computations to the four dimensional case one can obtain the generic expression
for the entropy,
\begin{equation}\label{LLEntropy}
    S-S_{0} = \beta Q_{\xi}|_{\partial \Sigma_{H}},
\end{equation}
where $S_{0}$ stands for higher order corrections to the value of entropy \cite{Aros:2001gz}. The
evaluation of Eq.(\ref{LLEntropy}) reproduces the results of \cite{Aros:2000ij}.

\section{Discussion and conclusions}

In this work it has been recovered the basic statistical mechanical relations of black holes in
the canonical ensemble using a Hamiltonian approach.

The two different boundary conditions presented, proper of $\Lambda=0$ and $\Lambda<0$
respectively, define a canonical ensemble. This might lead to think that somehow both boundary
conditions are equivalent. It is direct to prove, however, that this is not the case. For
$\Lambda<0$ the boundary condition at infinity let $\omega^{ab}$ undetermined because of
Eq.(\ref{BoundarytermNegative}), however since the limits $x\rightarrow \mathbb{R}\times \partial
\Sigma_{H}$ and $l\rightarrow 0$ do not commute then this boundary condition for $\Lambda<0$ can
not be extrapolated to the asymptotically flat case. The fundamental result of this work is that
there can be more than one set of boundary conditions that lead to the canonical ensemble. The
other result is to confirm that the horizon is a fundamental element in black holes
thermodynamics.

The analysis in this work probably can be easily extended to higher dimensions, except for one
important point, the phase variables $(e^{a}_{i},\pi_{a}^{i})$ are a feature proper of only four
dimensions\footnote{In three dimensions is not even necessary a projection like Eq.
(\ref{DecomposingTheMomentum})}. In higher dimensions there should be in principle more variables
in the phase space, $(e^{a}_{i}, \omega^{ab}_{i},\pi_{a}^{i},\pi_{ab}^{i})$, or the reduction to a
single pair of variables should be done in at least a different way.

\appendix

\section{Asymptotically flat versus ALAdS}\label{Naive}

To have an idea of how asymptotical flat spaces are not well behaved one can naively sketch
Eq.(\ref{PartitionFunction}) for four dimensional Schwarzschild solution. Here the entropy $S\sim
\pi r_{+}^2$ and the energy $E \sim r_{+}$, therefore the partition function
\begin{equation}\label{ZSch}
    \mathbf{Z}(\beta)_{Sch} \sim \int_{0}^{\infty} dr_{+} e^{-\beta r_{+} + \pi r^{2}_{+}}
\end{equation}
which is clearly divergent.

On the other hand with a negative cosmological constant the scenario changes radically, since for
Schwarzschild-AdS $E \sim r_{+}(1+l^{-2}r_{+}^{2})$, $l$ is the AdS radius, yielding the
completely different $\mathbf{Z}(\beta)$ function
\begin{equation}\label{ZSchNegative}
    \mathbf{Z}(\beta)_{Sch-AdS} \sim \int_{0}^{\infty} dr_{+} e^{-\beta (r_{+}+ r_{+}^{3}/l^{2}) + \pi r^{2}_{+}}
\end{equation}
which trivially converges.

\section{Kerr-AdS}\label{KerrAdS}
The Kerr-AdS geometry in Boyer-Lindquist-type coordinates can be expressed by the vierbein
\[
e^{0}=\frac{\sqrt{\Delta _{r}}}{\Xi \rho }(dt-a\sin ^{2}\theta d\varphi ),  \textrm{  } e^{1}=\rho
\frac{dr}{\sqrt{\Delta _{r}}},
\]
\begin{equation}\label{KerrTetrad}
e^{2}=\rho \frac{d\theta }{\sqrt{\Delta _{\theta }}}, \textrm{  } e^{3}=\frac{\sqrt{\Delta
_{\theta }}}{\Xi \rho }\sin \theta (adt-(r^{2}+a^{2})d\varphi ),
\end{equation}
with $\Delta _{r}=(r^{2}+a^{2})\left( 1+\frac{r^{2}}{l^{2}}\right) -2mr$, $%
\Delta _{\theta }=1-\frac{a^{2}}{l^{2}}\cos ^{2}\theta $, $\Xi =1-\frac{a^{2}}{l^{2}}$ and $\rho
^{2}=r^{2}+a^{2}\cos ^{2}\theta $.

This vierbein indeed has the form described in Eq.(\ref{ADMlabel}) and the resulting metric has
the required ADM form of Eq.(\ref{ADM}) as well.

In this coordinates the horizon is define by the largest zero of $\Delta_{r}$, called $r_{+}$. The
angular velocity of the horizon \cite{Hawking:1998kw} is
\[
\Omega = \frac{a}{r^2_{+}+a^2}.
\]
The mass and the angular momentum are found evaluating the charge (\ref{NoetherCharge}) for the
Killing vectors $\frac{\partial }{\partial t}$ and $\frac{\partial }{\partial \varphi }$,
respectively
\begin{equation}\label{MJKerr}
Q\left(\frac{\partial }{\partial t}\right)=\frac{m}{\Xi ^{2}}=M;  \textrm{ }Q\left(\frac{\partial
}{\partial \varphi }\right)=\frac{ma}{\Xi ^{2}}=J
\end{equation}
in agreement with Ref. \cite{Henneaux:1985tv}.

\section{Topological Black Holes}\label{TopologicalBHSec}
The restricted Lovelock gravities determined by the constants $\alpha_{p}$, in terms of $k<n-1$,
in Eq.(\ref{alphaCoeffients}) give rise to different topological black holes solutions depending
on $k$. Each one of them can be described by the vielbein
\begin{equation}
\label{vielbein} e^{0}=f(r) dt\textrm{  } e^{1}= \frac{1}{f(r)}dr\textrm{  }e^{m} = r
\tilde{e}^{m}\,,
\end{equation}
and its associated torsion free connection
\begin{equation}\label{spinconnection}
 \omega^{01}=\frac{1}{2}\frac{d}{dr}f(r)^{2} dt\textrm{  }\omega^{1m}= f(r)
\tilde{e}^{m}\textrm{  }\omega^{mn} =  \tilde{\omega}^{mn},
\end{equation}
where \begin{equation}\label{f2} f^{2}(r) = \gamma +\frac{r^{2}}{l^{2}}-\sigma \left(
\frac{C_{1}}{r^{d-2k-1}}\right)^{1/k},
\end{equation}$\sigma =(\pm 1)^{(k+1)}$, and the integration constant $C_{1}$ is identified as
\[
C_{1}=2G_{k}(M),
\]
where $M$ stands for the mass. $\tilde{e}^{m}=\tilde{e}^{m}_{i}(y)dy^{i}$ and
$\tilde{\omega}^{mn}$ are a vielbein and its associated torsion free connection on the transverse
section with $m=2\ldots d-1$. $\tilde{R}^{mn}=\gamma e^{m}e^{n}$. The $y^{i}$'s are an adequate
set of coordinates.

It is straightforward to prove that the mass can be obtained by evaluating
Eq.(\ref{NoetherCharge}) for the Killing vectors $\xi$ at $\partial \Sigma_{\infty}$,
\begin{equation}\label{MTBH}
Q\left(\frac{\partial }{\partial t}\right)=M.
\end{equation}

\section{Explicit expressions}\label{ExplicitExpressionsofConstr}

The different constraints $H_{\perp }$, $H_{i}$, and $J_{ab}$ can be written explicitly as
\begin{eqnarray*}
  \frac{1}{2}H_{\perp } &=& \eta ^{a}\partial _{i}\pi _{a}^{i}-\frac{1}{2}%
E_{d}^{s}\partial _{[l}e_{s]}^{d}\eta ^{b}\pi _{b}^{l} \\
   &-& G_{\perp ij}^{ab}\pi _{a}^{i}\pi _{b}^{j}-g^{3/2}G^{mnpq}\lambda _{mn}^{0}\lambda _{pq}^{0},
\end{eqnarray*}
\begin{widetext}
\begin{eqnarray}
N^{m}H_{m}= N^{m}\,\left[\frac{1}{2}\left(g^{-1}E_{d}^{s}\partial
_{i}e_{k}^{d}\varepsilon_{mls}\varepsilon ^{ijk}e_{j}^{b}-E_{d}^{s}\partial
_{[m}e_{s]}^{d}e_{l}^{b}+\eta _{d}\partial _{[m}e_{l]}^{d}\eta ^{b}\right)\pi
_{b}^{l}\right.\nonumber \\
+ \left. e_{m}^{a}\partial _{i}\pi _{a}^{i}+G_{mij}^{ab}\pi _{a}^{i}\pi _{b}^{j}+%
\frac{1}{2}N^{(m}e_{i}^{a}e_{j}^{b} J_{ab}\epsilon^{ijn)}\lambda _{mn}^{0} \right]\\
  -N^{i}\omega _{i}^{ab}\ J_{ab}\nonumber,
\end{eqnarray}
\end{widetext}
and
\begin{equation}
J_{ab}=2\varepsilon _{abcd}\frac{\partial e_{j}^{c}}{\partial x^{i}}%
e_{k}^{d}\varepsilon ^{ijk}-\frac{1}{2}(\pi _{a}^{i}e_{bi}-\pi _{b}^{i}e_{ai}),
\end{equation}
where
\begin{equation}
\lambda _{pq}^{0}=\frac{1}{2g}G_{pqmn}E_{a}^{(m}\partial _{i}e_{j}^{a}\epsilon ^{ijn)},
\label{lambda0}
\end{equation}
\begin{equation}
G_{\perp ij}^{ab}=\frac{1}{16\sqrt{g}}[%
e_{i}^{a}e_{j}^{b}-2e_{j}^{a}e_{i}^{b}-g_{ij}\eta ^{a}\eta ^{b}],
\end{equation}
and
\begin{equation}
G_{mij}^{ab}=\frac{1}{16\sqrt{g}}[g_{ij}\eta ^{a}e_{m}^{b}+2g_{im}(e_{j}^{a}\eta
^{b}-e_{j}^{b}\eta ^{a})].
\end{equation}
\section{Microcanonical boundary term}

To transform the action $\tilde{I}_{EH}$ into the microcanonical ensemble action is necessary to
add a boundary term that change the boundary conditions from $\delta N=0$ to a fixed energy
density $e$ at the boundary. This is simply achieved by subtracting from
Eq.(\ref{ImprovedDecomposed}) the term
\[\int_{\partial \Sigma} ( e N - V^{m}p_{m})\, d^{2}\sigma, \]
which is the boundary term $B$. This result leads to the new the action principle
\begin{equation}\label{Microcanonicalaction}
\hat{I}_{EH} = \int_{\mathcal{M}} (\dot e^{a}_{i} \pi_{a}^{i}+ N H_{\perp}+ N^{i} H_{i} +
\omega^{ab}_{t} J_{ab})\, dt\wedge d^{3}x,
\end{equation}
which should be suitable for the microcanonical ensemble. Unfortunately the analysis of the
thermodynamics in this case is not straightforward, and will be discussed elsewhere.

\begin{acknowledgments}
I would like to thank Abdus Salam International Centre for Theoretical Physics (ICTP) for the
associate award granted. This work was partially funded by grants FONDECYT 1040202 and DI 06-04.
(UNAB). Part of this work was written at ICTP.
\end{acknowledgments}

\vspace{0.3in}
\providecommand{\href}[2]{#2}\begingroup\raggedright\endgroup
\end{document}